# Magnetic Properties of $RuSr_2RECu_2O_8$ (RE=Gd, Eu) and $Ru_{1-x}Sr_2GdCu_{2+x}O_{8-y}$ Superconductors

P.W. Klamut*, B. Dabrowski, S.M. Mini, M. Maxwell, S. Kolesnik, M. Mais

Department of Physics, Northern Illinois University, DeKalb, IL 60115

A. Shengelaya, R. Khasanov, I. Savic, H. Keller

Physik Institut der Universität Zürich, Switzerland

T. Graber, J. Gebhardt, P.J. Viccaro

The Center for Advanced Radiation Sources, The University of Chicago

Y. Xiao

Southern Illinois University at Carbondale

**Abstract**

Synthesis in 1% oxygen at 930°C leads to non-superconducting $RuSr_2GdCu_2O_8$ in which superconductivity subsequently can be induced with oxygen annealings without observable stoichiometric changes. Detailed *ac* susceptibility and zero-field muon spin rotation experiments reveal that the superconducting compounds always exhibit a lower temperature of the magnetic transition and larger values of the internal magnetic field in the ordered state of the Ru sublattice. The same annealings in oxygen, however, lead to non-superconducting $RuSr_2EuCu_2O_8$.

The magnetic properties of a newly discovered series of $Ru_{1-x}Sr_2GdCu_{2+x}O_{8-y}$ superconductors (maximum $T_c$=72 K for x=0.4) are reported. Low temperature behavior of the magnetization shows a significant contribution of the paramagnetic system of Gd ions to the magnetic response below the superconducting transition. Muon-spin rotation experiments reveal the presence of bulk magnetic transitions at low temperature for compounds with x≤0.4. XANES Cu-K absorption edge data show an increase in Cu valence with x.

___________

*corresponding author, email: klamut@niu.edu

**Introduction**

Recent reports of the apparent coexistence of superconductivity (SC) and weak-ferromagnetism (FM) in ruthenocuprates [1,2] have triggered intense interest in the properties of these materials. The compounds that exhibit this unusual behavior are $RuSr_2RECu_2O_8$ (Ru-1212) [3,2] and $RuSr_2(RE_{2-x}Ce_x)Cu_2O_{10-y}$ (Ru-1222) (RE=Gd, Eu) [1] and they belong to the family of high temperature superconductors. The crystal structure of Ru-1212 can be described based on its similarity to $REBa_2Cu_3O_{7-y}$ (RE123) superconductors. The structure contains double $CuO_2$ planes separated by a single oxygen-less Gd layer. The Ru atoms, coordinating full octhaedra of oxygens, form the $RuO_2$ planes, which replace the Cu-O chains present in RE123. Recent neutron diffraction experiments have shown that the dominant magnetic interactions present in $RuSr_2RECu_2O_8$ are of the antiferromagnetic (AFM) type with Ru moments forming the G-type antiferromagnetic structure [4, 5]. The ferromagnetism observed in these compounds was proposed to originate from the canting of Ru moments that give a net moment perpendicular to the c-axis [4, 5]. This description is similar to that suggested for $Gd_2CuO_4$, a non-superconducting weak ferromagnet, where the distortions present in the $CuO_2$ plane permit the presence of the antisymmetric Dzialoshinski-Moriya superexchange interactions in the system of Cu magnetic moments [6].

Reports of both superconducting [2,3] and non-superconducting samples [7, 8] of Ru-1212 create a need for a detailed experimental approach aimed at determining the factors which influence the superconductivity of this material. In the first part of the paper we present the results of a modified synthesis technique for Ru-1212 which allows superconductivity to be induced in a controlled manner.

The second part presents select magnetic properties of a new series of superconducting compounds, with the formula $Ru_{1-x}Sr_2GdCu_{2+x}O_{8-y}$, that have been successfully synthesized recently in high pressure oxygen [9]. Investigating the peculiar properties of these materials could lead to a better understanding of SC and FM coexistence observed in parent $RuSr_2RECu_2O_8$.

**Experimental**

The crystal structure of all reported samples was examined by the X-ray powder diffraction method using a Rigaku Inc. X-ray Diffractometer (CuKα radiation).

The *ac* susceptibility ($\chi'$) and *dc* magnetization were measured using a Quantum Design Physical Properties Measurement System. In the *ac* measurement data was collected upon warming from a zero-field-cooled state at *ac* field of 1 Oe (f= 200 Hz). The µSR measurements were performed on beam line πM3 at the Paul Scherrer Institute (PSI, Switzerland) using low momentum muons (29MeV/c). XANES measurements were carried out at the Chem.Mat. CARS undulator beamline at the Advanced Photon Source in Argonne.

**Results**

1. $RuSr_2GdCu_2O_8$ vs. $RuSr_2EuCu_2O_8$

Polycrystalline samples of $RuSr_2RECu_2O_8$ (RE=Gd, Eu) were synthesized by solid state reaction of stoichiometric oxides of $RuO_2$, $Eu_2O_3$, $Gd_2O_3$, CuO and $SrCO_3$. After calcination in air at 920°C, the material was ground, pressed into pellets and annealed at 930°C in the flow of 1% of oxygen in argon. Several annealings, with intermediate grinding and pelletizing, performed in this atmosphere were always followed by cooling in Ar in order to avoid the reformation of $SrRuO_3$ impurities. The repeated procedure resulted in single phased material with no traces of impurities seen in the powder X-ray diffraction patterns. This modified method of synthesis always leads to non-superconducting samples of both Eu and Gd based compounds.

Figures 1 presents the temperature dependencies of the ac susceptibility for non-superconducting $RuSr_2GdCu_2O_8$ and two other chunks of this material after subsequent annealing at 1060 C in flowing oxygen for 24 and 180 h, respectively. The oxygen annealing gradually induces superconductivity in the material. The 180 h annealed sample already shows the saturated effect. The low temperature diamagnetic response corresponds to the diamagnetic shielding of the whole volume. The thermogravimetry analysis performed in flowing oxygen for the 1060 C annealing did not show any observable changes in the oxygen concentration. Thus, ordering of the Gd/Sr and Ru/Cu atoms within the structure or changes in the coherence of structural distortions can be considered as a candidate effect of the oxygen annealing and need to be investigated by means of detailed neutron diffraction experiments. Our preliminary XANES Cu-K absorption edge measurements performed at room temperature show virtually no difference (<S>=5.78 eV) between superconducting and non-superconducting samples suggesting the same oxidation state of Cu. Detailed XANES measurements of the temperature dependencies of the valence state of Ru and Cu are in progress and will be communicated separately.

Oxygen annealing of non-superconducting $RuSr_2GdCu_2O_8$ was also found to lower the temperature of the magnetic transition of the Ru sublattice ($T_m$=130 vs. 136 K - see inset in Fig. 1). The real part of ac susceptibility below the magnetic transition always *increases* after oxygen annealing. This suggests enhancement of the weak-ferromagnetic component in the magnetically ordered sublattice of Ru. Fig. 2 shows the results of zero field muon spin rotation spectroscopy measurements that provide a direct measure of the internal magnetic field. Open circles present the data for non-superconducting $RuSr_2GdCu_2O_8$ synthesized according to the procedure described above. Comparing it with data communicated by Bernhard et al in [2] for the superconducting material (closed circles) reveals that the temperature of the magnetic transition is lowered by approx. 10 K and the internal magnetic field remains larger by approx. 10% for a superconducting sample. These results indicate an interesting correlation between the superconducting and magnetic properties of $RuSr_2GdCu_2O_8$.

Fig. 3 shows the temperature dependencies of *ac* susceptibility for three samples of $RuSr_2EuCu_2O_8$ that underwent the same thermal treatments in $1\%O_2/Ar$ and oxygen as $RuSr_2GdCu_2O_8$ presented in Fig.1. These annealings, while also slightly decreasing the temperature of the magnetic transition (see inset in Fig.3), do not lead to superconducting material. Note that the susceptibility of $RuSr_2EuCu_2O_8$ in its ordered state remains significantly smaller than for Gd based compounds. The paramagnetic contribution of the $Gd^{3+}$ sublattice (compare the ac susceptibility dependence for non-superconducting $Gd^{3+}Ba_2Cu_3O_{6.2}$ presented with solid line in Fig. 1) remains too small to fully account for this difference. Thus, one can conclude that the weak-ferromagnetic component in $RuSr_2GdCu_2O_8$ is enhanced compared to its Eu-based analogue. This result points toward the role of $Gd^{3+}$ magnetic moments possibly enhancing the weak-ferromagnetism of the Ru sublattice or better structural order and larger distortions for $RuSr_2GdCu_2O_8$. Superconducting $RuSr_2EuCu_2O_8$ can still be obtained, but our experiments indicate this occurs only in the presence of a small fraction of secondary phases (predominantly $SrRuO_3$). The solid line in Fig.3 shows the onset of the superconducting transition for such material. Because the presence of secondary phases creates a favorable situation for the formation of structural defects, the superconductivity in this sample should be considered within the scope of our recent finding that partial Cu->Ru substitution in $Ru_{1-x}Sr_2GdCu_{2+x}O_{8-y}$ leads to a significant increase of the superconducting $T_c$ (see later in the text). Research aimed toward synthesizing Eu based analogues of these compounds is now in progress.

2. $Ru_{1-x}Sr_2GdCu_{2+x}O_{8-y}$

Polycrystalline samples of $Ru_{1-x}Sr_2GdCu_{2+x}O_{8-y}$ (x=0, 0.1, 0.2, 0.3, 0.4, 0.75) were prepared by the solid-state reaction of stoichiometric $RuO_2$, $SrCO_3$, $Gd_2O_3$ and CuO. After calcination in air at 920°C the samples were ground, pressed into pellets, and annealed at 970°C in flowing oxygen. The samples were sintered at 1060 and 1085°C in a high-pressure oxygen atmosphere (600 bar). The Ru-1212-type structure formed for all compositions with traces of other impurity phases present (predominantly $SrRuO_3$). The decrease of both *a* and *c* lattice parameters with x can indicate increased hole doping [9].

Fig. 4 presents the temperatures of the onsets of resistive superconducting transitions for the series. The maximum $T_c$ was found at 72 K for the x=0.4 composition (see [9] for details). Fig. 5 presents the temperature dependencies of the ac susceptibility for x=0, 0.1, 0.3 and 0.75 compounds. $T_{c1}$ marks the onsets of the superconducting transitions. The magnetic ordering of the Ru sublattice, present below $T_M$=130 K for parent $RuSr_2GdCu_2O_8$, can be traced to only slight irreversibility of FC and ZFC branches of magnetization observed below 120 K and 100 K for x=0.1 and 0.2 samples [9]. Muon spin rotation experiments performed for the x=0.1 sample shows a slight increase of the relaxation rate below 120 K. However, this change can not be attributed to the bulk response of the material. Thus, we could conclude that this observed feature originates from a compositional inhomogeneity, for example the formation of Ru rich clusters in the Ru/Cu-O planes.

The high magnetic field magnetization data collected in 1000 Oe steps at 4.5 K for the $Ru_{1-x}Sr_2GdCu_{2+x}O_{8-y}$ series is presented in Fig.6. The magnetization for several x≠0 samples, shown with solid lines, appears to converge to the dependence obtained for non-superconducting $GdBa_2Cu_3O_{6.2}$ (open circles). The closed circles represent the magnetization of the $RuSr_2GdCu_2O_8$. By comparing the parent and x≠0 samples it can be concluded that no additional contribution from the Ru sublattice to the measured signal is observed for the diluted Ru sublattice, i.e. magnetic response is characteristic of paramagnetic $Gd^{3+}$ ions as seen in $GdBa_2Cu_3O_{6.2}$. We should note that the extra contribution of magnetization observed for $RuSr_2GdCu_2O_8$ suggests considerable ferromagnetic alignment of the Ru moments at high magnetic fields (increase by about 1 $\mu_B$).

Fig. 7 presents the M(H) dependencies measured for the x= 0, 0.3 and 0.75 samples at 4.5, 20, and 50 K and small magnetic fields changed between -500 Oe and 500 Oe. The hysteresis loops can be interpreted as the superposition of the magnetic and superconducting components. For x≠0 samples remnant magnetization is observed only below the superconducting transition. The first penetration fields at T=4.5 K, (see the local minima for corresponding virgin parts of M(H) loops), are 5, 8 and 14 Oe for x=0, 0.3 and 0.75, respectively. Larger positive contributions to the low temperature magnetization are observed for samples with smaller x. The field dependent reentrant behavior of the magnetization observed at low temperatures for the whole series (see inset in Fig.7 for x=0.75 sample) suggests that this contribution originates in the paramagnetic response of the Gd sublattice. The constrained dimensionality of the superconducting phase, possibly evolving along the series, may be related to the observed behavior [9].

Fig. 8 presents the results of zero-field muon spin rotation measurements for x=0.1, 0.3 and 0.4 compositions. The asymetry parameters measured in this experiment allow us to conclude that we observe the bulk magnetic transitions resulting in the presence of an internal magnetic field below 13, 6 and 2 K, for x=0.1, 0.3 and 0.4 respectively. Cu or Ru magnetic sublattices are the only candidate systems that can be responsible for this behavior. Preliminary XANES data collected at room temperature show that the estimated characteristic energies (proportional to the effective charge per edge atom) for Cu K-absorption spectra are S=5.78, 5.97, 6.08 and 6.13 eV for x=0, 0.15, 0.3 and 0.5 compositions. Comparing these values to S=6.19 and 6.45 eV measured for x=0 and 0.5 samples of YBa2Cu3O6+x [10], can suggest substantially lower hole doping per Cu atom for compounds investigated here.

**Concluding remarks**

The modified synthesis conditions found for $RuSr_2GdCu_2O_8$ lead us to obtain both superconducting and non-superconducting compounds of seemingly the same composition. Comparison of the magnetic properties of these samples, utilizing the dc magnetization, ac susceptibility and μSR data, reveal that the weak ferromagnetic component is always enhanced for the superconducting compounds, whereas the temperature of the magnetic ordering is slightly lowered (130 vs. 136 K). The effect can be preliminary attributed to the different level of disorder or structural distortions present in the crystal structure, which in turn would influence the weak ferromagnetic response observed for the antiferromagneticaly ordered and canted sublattice of the Ru moments. Detailed neutron diffraction experiments on $Gd^{160}$ enriched compound are planned to further elucidate the role of the structure disorder in determining the properties of this material. The XANES Cu-K absorption spectra show virtually no difference between non- and superconducting samples suggesting the same Cu valence state for both compounds. This prompts the unresolved question on the nature of possible interdependence between superconducting and weak-ferromagnetic behavior. Further experiments, centered on the determination of the spatial origin of the superconducting phase as well as the peculiarities in the interlayer charge distribution for $RuSr_2GdCu_2O_8$, are clearly required in order to explain observed behavior.

The magnetic properties of the novel series of $Ru_{1-x}Sr_2GdCu_{2+x}O_{8-y}$ superconductors (maximum $T_c$=72 K for x=0.4) reveal the dominant contribution of the paramagnetic response of the $Gd^{3+}$ sublattice at low temperatures. These results can be interpreted as indicative of the constrained dimensionality of the superconducting phase that apparently evolve along the series toward quasi-two dimensional behavior characteristic for the x=0 parent compound. The low temperature magnetic transitions observed in zero-field µSR experiment for x<0.4 compositions should be attributed to the magnetic response of the Ru/Cu-diluted sublattice or Cu moments in $CuO_2$ planes. Considering the latter, and assuming spatial homogeneity of the superconducting phase, it would create challenging consequences in understanding the origin of the superconducting condensate in these compounds. However, we should note that the µSR data can be also interpreted as indicative of a microscopic coexistence of the AF and SC order parameter as it has been observed for other underdoped systems: see [11] for the AF correlations observed for underdoped $La_{2-x}Sr_xCuO_4$ and $Y_{1-x}Ca_xBa_2Cu_3O_6$ for which the microscopic inhomogeneity of the charge distribution in $CuO_2$ planes and associated stripes formation have been proposed [12]. Further extensive investigations of the peculiar properties of these new compounds are now in progress.

**Acknowledgement**

Work supported by the ARPA/ONR and by the State of Illinois under HECA.
P.W.K., A.S., R.K. and I.S. would like to thank Dr. D. Herlach and U. Zimmerman of PSI, Villigen for their valuable assistance with the μSR experiments. S.M.M., T.G., J.G., Y.X., and P.J.V. gratefully acknowledge support by the National Science Foundation (CHE-9871246 and CHE-9522232) and the use of the Advanced Photon Source was supported by the U.S. Department of Energy, Basic Energy Sciences, Office of Science, under Contract No. W-31-109-Eng-38.

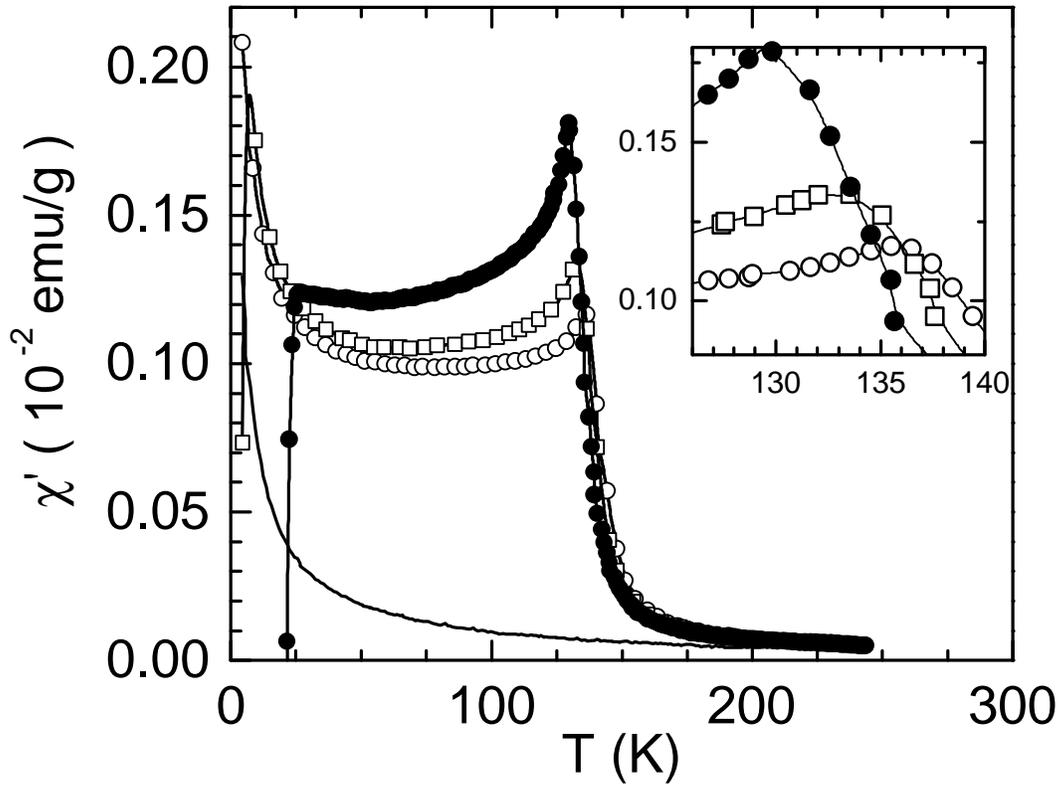

**Fig.1.** Temperature dependencies of *ac* magnetic susceptibility of $RuSr_2GdCu_2O_8$ after different thermal treatments (only positive part shown): open circles - as synthesized sample (930°C in 1% of oxygen); open squares and closed circles: the same material after additional annealings in oxygen at 1060°C for 24 and 180 h, respectively. Inset shows the corresponding shift of $T_m$ (see text). The solid line represents the susceptibility for non-superconducting $GdBa_2Cu_3O_{6.2}$. $H_{ac}$=1 Oe, f=200 Hz.

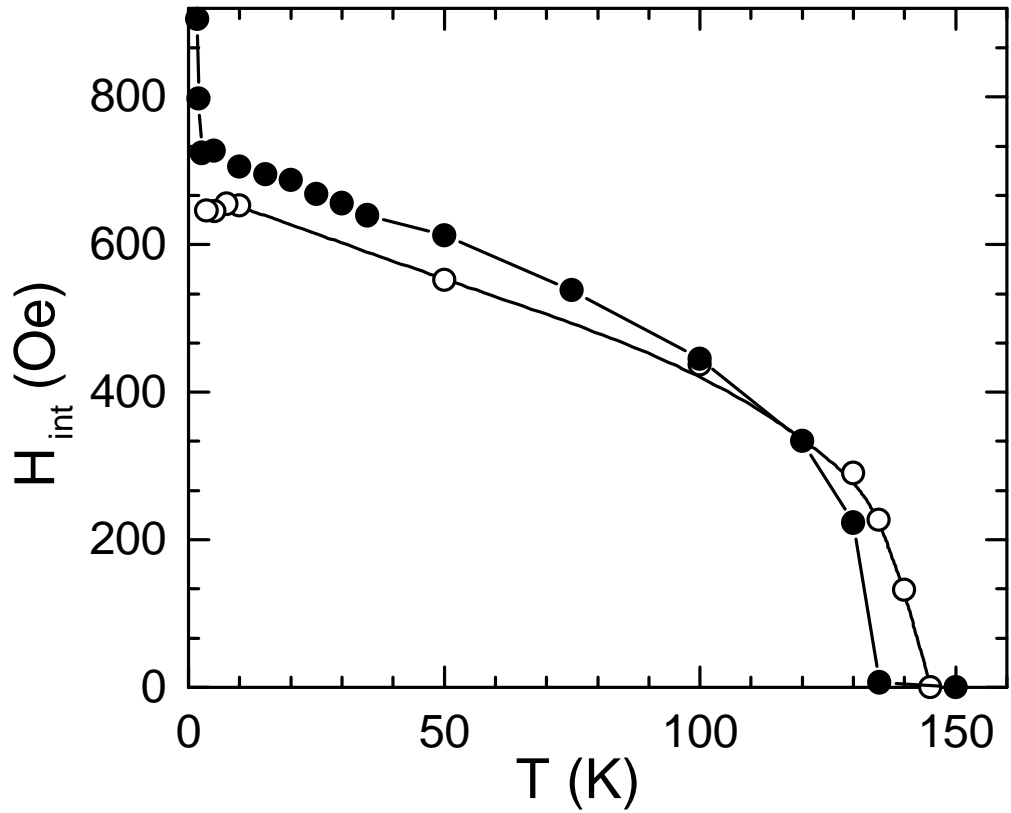

**Fig.2.** Temperature dependencies of the internal magnetic field for $RuSr_2GdCu_2O_8$ calculated from the results of zero-field muon spin rotation spectroscopy measurements.
Open circles: data for here reported non-superconducting sample. Closed circles: values calculated from the results presented in [2] for superconducting material.

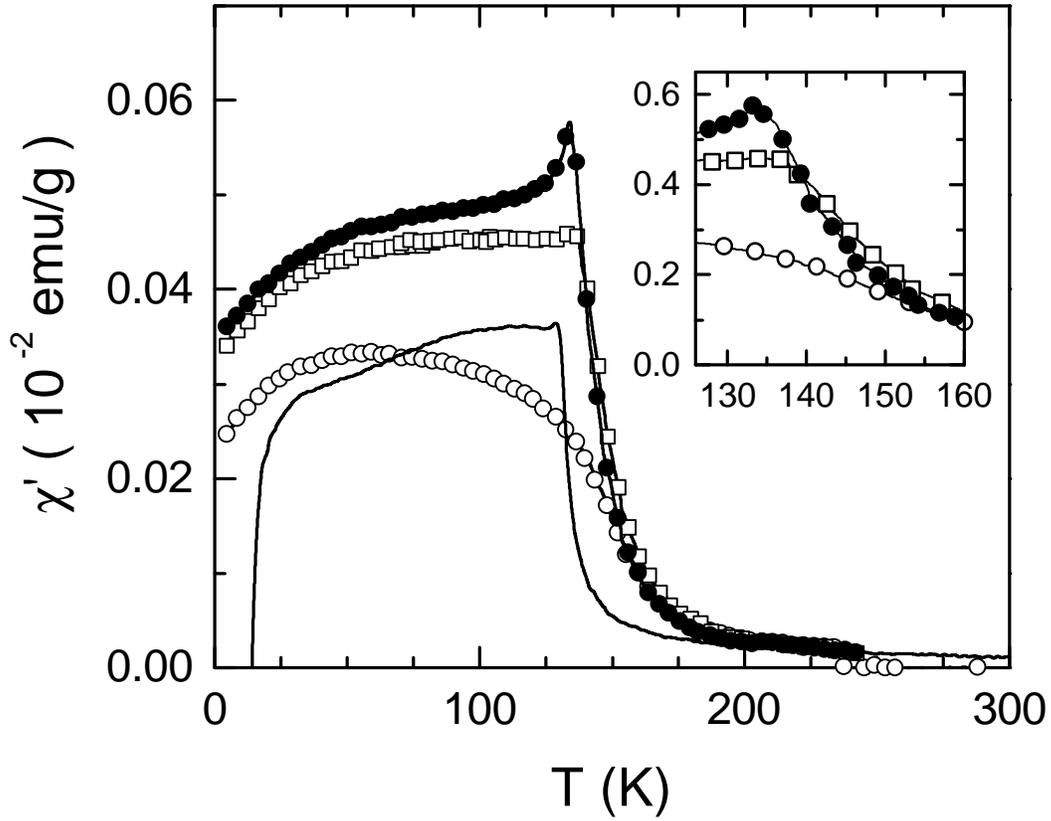

**Fig.3.** Temperature dependencies of *ac* magnetic susceptibility of $RuSr_2EuCu_2O_8$ after several annealings (only positive part shown) - notation as on Fig.1. The solid line represents the superconducting $RuSr_2EuCu_2O_8$ sample with a small amount of secondary phases present. Inset shows the shift of $T_m$. $H_{ac}$=1 Oe, f=200 Hz.

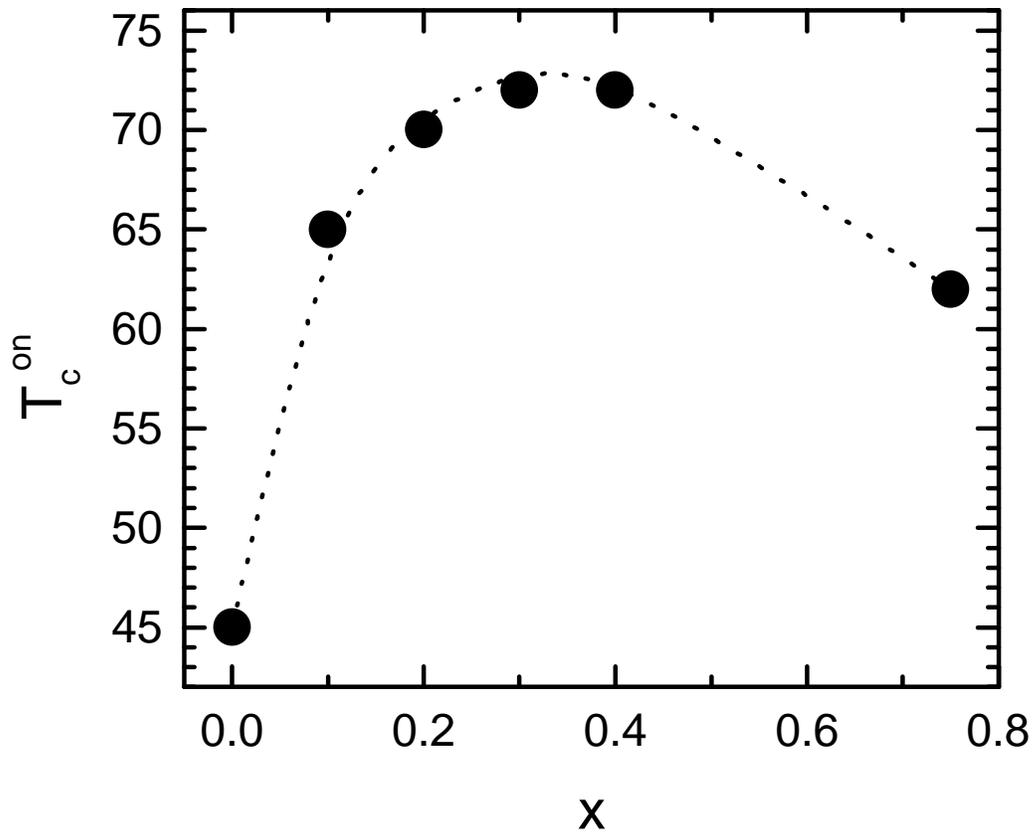

**Fig.4.** $T_c^{on}$ versus x for $Ru_{1-x}Sr_2GdCu_{2+x}O_{8-y}$ series.

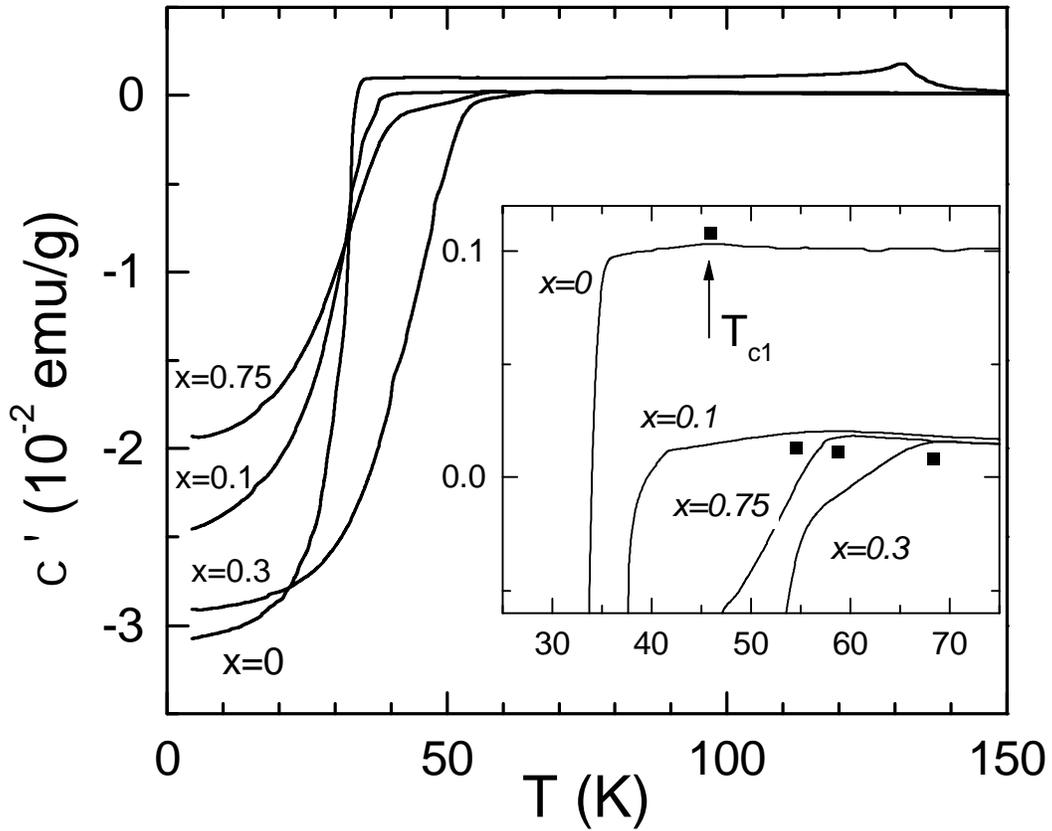

**Fig.5.** Temperature dependencies of the *ac* susceptibility for x=0, 0.1, 0.3 and 0.75 compositions of $Ru_{1-x}Sr_2GdCu_{2+x}O_{8-y}$. Insets present in the expanded scale the onsets of the superconducting transitions ($T_{c1}$) marked with squares.

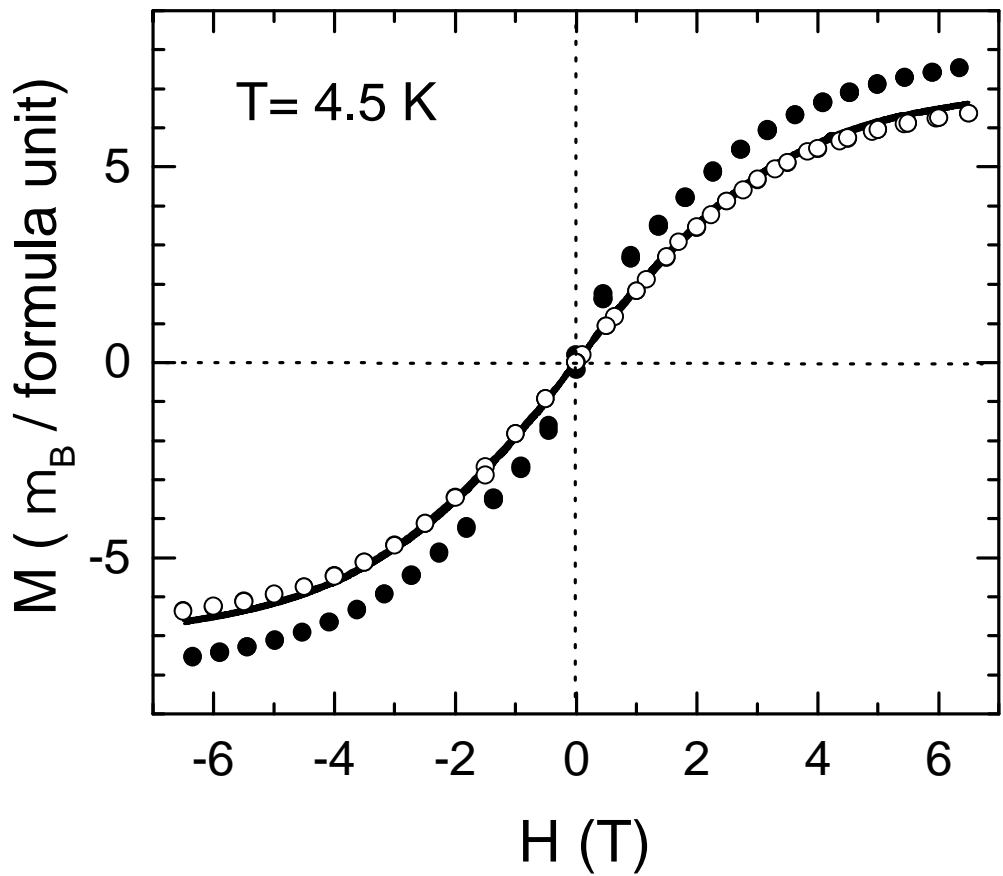

**Fig.6.** The magnetic field dependencies of the magnetization measured at 4.5 K for Ru$_{1-x}$Sr$_2$GdCu$_{2+x}$O$_{8-y}$ samples (solid lines converge to one curve). Closed circles show the behavior of superconducting RuSr$_2$GdCu$_2$O$_8$, open circles of non-superconducting GdBa$_2$Cu$_3$O$_{6.2}$.

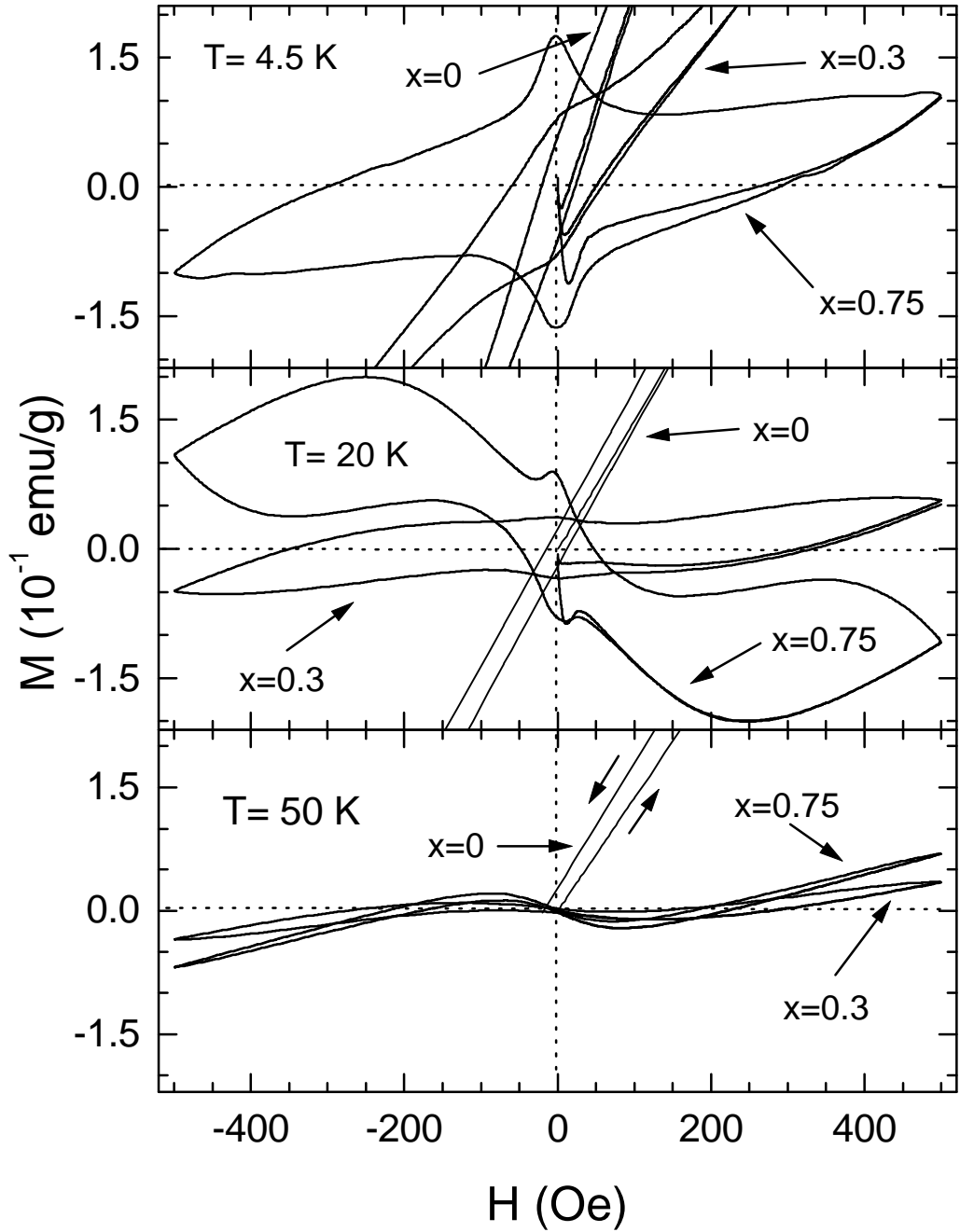

**Fig.7.** The magnetic field dependencies of the magnetization measured at 4.5, 20 and 50 K for x=0, 0.3 and 0.75 samples. The field cycled between -500 and 500 Oe. Inset shows the temperature dependence of the zero-field cooled (ZFC) and field cooled (FC) magnetization ($H_{dc}$=500 Oe) for x=0.75 sample, dotted line present corresponding dependence for $GdBa_2Cu_3O_{6.2}$.

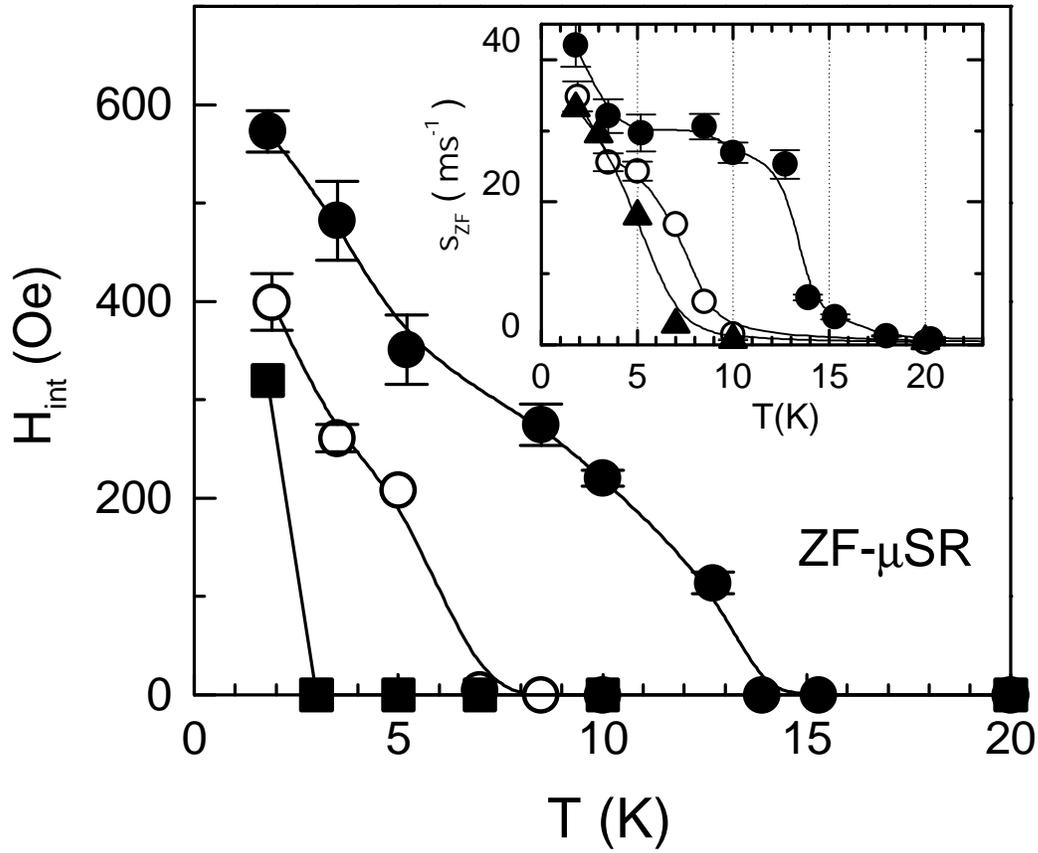

**Fig.8.** Temperature dependencies of the internal magnetic field for $Ru_{1-x}Sr_2GdCu_{2+x}O_{8-y}$ measured in zero-field muon spin rotation experiment. Samples: x=0.1 (closed circles), 0.3 (open circles), and 0.4 (closed squares). Inset shows the corresponding temperature dependencies of the relaxation rates σ.